*Research Article*

# Secured Approach towards Reactive Routing Protocols Using Triple Factor in Mobile Ad Hoc Networks

**Mohammad Riyaz Belgaum**[1, *]**, Shahrulniza Musa**[1]**, MazlihamMohd Su'ud**[1]**,
Muhammad Alam**[1,2]**, Safeeullah Soomro**[3] **and Zainab Alansari**[3,4]

[1]Malaysian Institute of Information Technology, Universiti Kuala Lumpur, Malaysia.
mohammad.riyaz@s.unikl.edu.my, (shahrulniza‖mazliham‖mansoor)@unikl.edu.my
[2]Ilma University, Karachi, Pakistan
m.alam@ilmauniversity.edu.pk
[3]College of Computer Studies, AMA International University, Kingdom of Bahrain.
(s.soomro‖zeinab)@amaiu.edu.bh
[4]University of Malaya, Kuala Lumpur, Malaysia.
z.alansari@siswa.um.edu.my
**\*Correspondence:** mohammad.riyaz@s.unikl.edu.my

**Abstract:** Routing protocols are used to transmit the packets from the source to the destination node in mobile ad hoc networks. The intruders seek chance to pierce into the network and becomes a cause of malfunctioning in the network. These protocols are always prone to attacks. During the phases of routing in different types of protocols, each of the attack finds a way to degrade the performance of the routing protocols. The reactive routing protocols DSR and AODV have lot of similar features and so are considered in this study. In order to transmit the packets safely, a secured approach using triple factor has been proposed. This triple factor computes the trust by using the direct information then verifies the reputation by collecting the information from the neighbouring nodes called distributed factor and uses cryptographic algorithm to ensure security. And to ensure that there are routes available to perform the routing process, the reasons for such attacks are studied so as to re-integrate back the nodes in to the network, once it has repented for being malicious before. The availability of routes increases the throughput.

## 1. Introduction

A set of mobile nodes that perform basic networking functions in an infrastructure less environment is said be a mobile ad hoc network (MANET), as demonstrated in figure 1. Nodes that fall within the communication range communicate with each other and which don't come in the range follow the concept of multi-hop for communication. In the network each node plays a dual role as a host by the forwarding and as a router in routing packets to the destination.

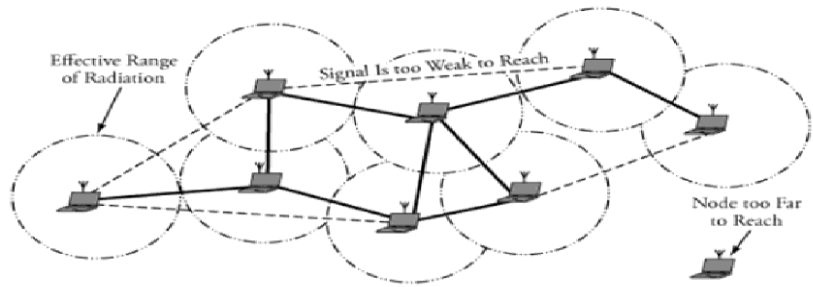

**Figure 1.** Mobile Ad Hoc Networks (MANET)

Maintaining security is an important function of any of the routing protocol in each phase of the networking function [1]. Because of the non-static topological behaviour of the network and due to being the network open which allows the network to grow and shrink due to addition and deletion of the nodes anytime gives chance for the intruder nodes to disturb the normal routing process. And if there does not exist a common regulatory authority for authenticating and guaranteeing the nodes then a reliable transmission is not possible.

In the current era of Internet of things (IoT), the wireless sensor networks functionality is similar to MANETS as both are dynamic and self-organized. From the figure 2, we can see that the IoT devices form into clusters and transmit the information through the network.

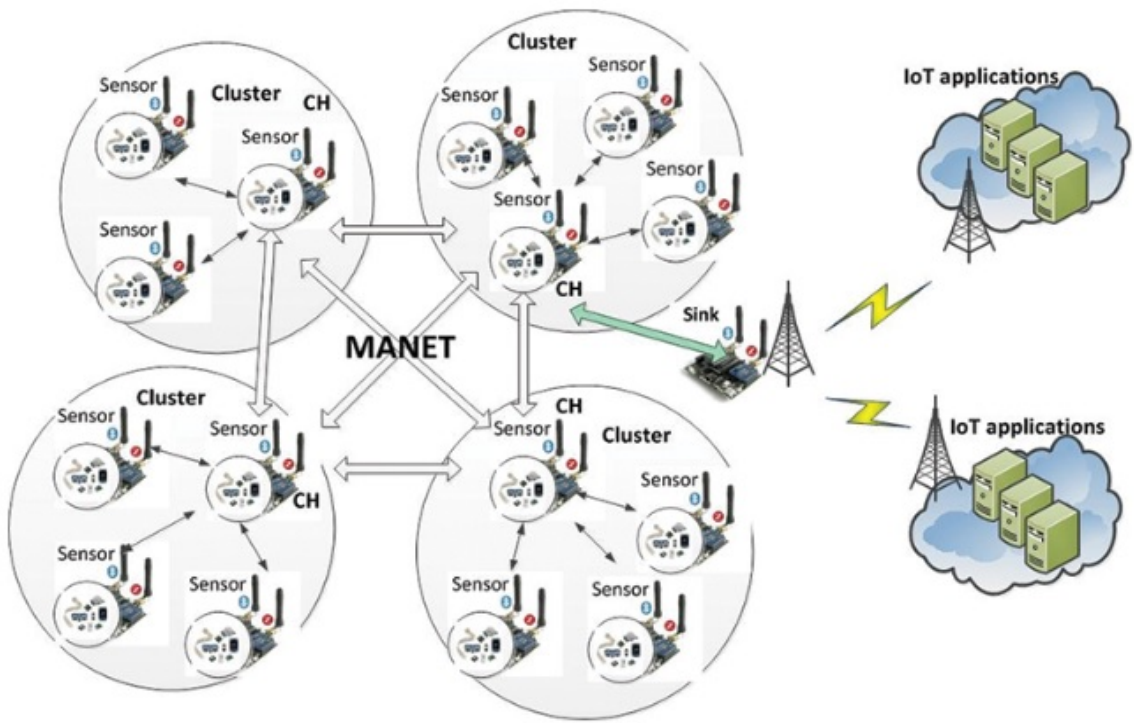

**Figure 2.** MANET-IoT network [2].

**1.1. Security Related Goals and Challenges**

Security services are needed to make sure that the data is transferred over the network with reliability and also the keeping the resources of the system protected. To attain the objectives, the categorizations of security services are: availability, confidentiality, authentication, integrity and non-repudiation [3, 4].

- **Availability:** Though the system is suffering from various problems like with bandwidth, connectivity but the availability service ensures that still the resources are available in a timely manner. The harmful effects of availability of a network are resource depletion attacks and packets dropping ratio.
- **Confidentiality:** The information prevailing in the network is not to be shared among all unauthorized nodes and this is achieved by Confidentiality. In order to achieve Confidentiality many encryption techniques can be used to make only the authorized nodes can share the transmission of information and the private and public keys.
- **Authenticity:** To prove a node as a legitimate user the network service used is Authenticity. The absence of this service can make any node in the network impersonate any node, and then having a total control capture and control over the complete network.
- **Integrity:** The data which is been transmitted in the network can be modified either wantedly or sometimes unwantedly. The Integrity network service ensures that the information which is been transmitted is not modified.
- **Non-repudiation:** This service guarantees that the message transmission has been done between the two parties and it cannot be denied. Also using this service it helps in detecting and isolating of compromised nodes in the network.

Communicating through the network in safe and secure way has been a challenging task because of

- Not being a stable infrastructure.
- The links in the network are prone to break and not secure.
- Scarcity or overload on the system resources
- The network topology being dynamic

In this study the plan of the research is to study the various reactive routing protocols in MANETS and analyze the threats and types of attacks in the routing protocols. The reasons for security threats are studied for giving a solution to meet the challenges of security in the network and carry out regular network operations in a secured way. The proposed approach will be used to enhance the existing reactive routing protocols by considering triple factor to improve security in while the network functions are carried out. Specifically the following issues will be addressed.

1. What are the different security threats for the reactive routing protocols?
2. What are the reasons for threats?
3. Strategies to make the network strong and secured

The researcher concentrates on the reactive routing protocols. The threats on these routing protocols are studied; as a result of security such malicious node will be deleted from taking part in forwarding the packet. Now the reasons for these types of threats are studied and the problems in this strategy are considered for research. The proposed architecture in MANETS will improve security by embedding triple factor in the reactive routing protocols while forwarding the packets from source to destination. This could serve as reference for other researchers to enhance other category of routing protocols based on their behavior to improve security in MANETS.

The study of working strategy of each of the reactive routing protocol along with the attacks on them is studied. Different protocols adopt different strategies when they are prone to attacks on the protocols. A study of reasons for the attacks and threats will be conducted which makes to adopt the triple factor to enhance security in the discovery and route maintenance during the process of sending the packets to the destination.

**2. Literature Review**

The following is the summary of various routing protocols based on their behavior designed for MANETS [5-9]. These protocols can be categorized as follows.

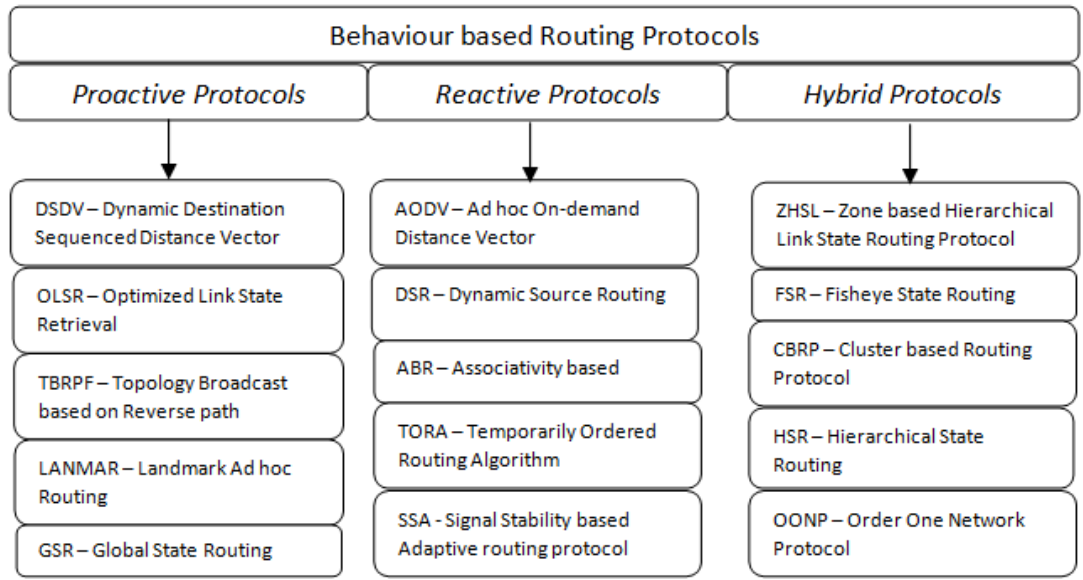

**Figure 3.** Routing Protocols

The researcher carries the research with the study of reactive routing protocols and the attacks on them. The working functionalities of each of these reactive routing protocols [10] are summarized as follows.

**2.1. Reactive Routing Protocols**

2.1.1. Dynamic Source Routing Protocol

The DSR protocol communicates by following two phases namely route discovery and maintenance [11]. The routing information is stored while the packets are forwarded. When a packet arrives at a node, it first checks its cache to ensure that the route for the destination node is available as it maintains the information of the recently used routes. When there are multiple routes to the destination then a shortest route with less hop count is selected. Because of the dynamic changes in the topology, there is a chance of routes being broken in the route maintenance phase still it ensures that the packet is safely transmitted to the target. There are two types of packets floating between source and destination as route request (RREQ) and route reply (RREP).

2.1.2. Ad Hoc on Demand Distance Vector Protocol

The functionality of AODV protocol is explained in [12]. The authors here proposed an new protocol using AODV as the base protocol where a fitness function is used. The traditional AODV protocol has a single path from the source to the destination node while in the proposed protocol, the authors used multipath. It is stated that the features of both DSR and DSDV are combined. The author explains the working of AODV protocol with two phases in them as route discovery and route maintenance. A method to identify the malicious node was explained in order to avoid forwarding of the information to the malicious node in the routing table. The solution given did not impose any overhead on the nodes in the network.

2.1.3. Temporally Ordered Routing Algorithm

The Temporally ordered Routing Algorithm considers the link reversal concept. This protocol doesn't allow the loops to occur [13]. There are three phases in this protocol as: (a) Route creation happens in first phase, (b) maintenance of route happens in second phase and (c) the elimination of invalid routes happen in third phase. All these phases go in a serial so as to safely transmit the packets from source to destination.

2.1.4. Associativity Based Routing

The Associativity Based Routing (ABR) protocol is free from loops and has no similar packets. Also no deadlock occurs in this protocol [14]. It focuses on route longevity. As there are very few broken communication links and less need for reconstruction of the routes the overhead involved is less. An improved version of ABR was to optimize the bandwidth and demand to reduce the overhead based on the position information was proposed. It was concluded that the path setup time was long for the routes which gave a scope for the future research to improve the ABR Protocol.

2.1.5. Signal Stability-based Adaptive Routing Protocol

The working of SSR routing protocol states that the large routing tables are not required for routing [15]. The network will not be congested with the control messages. From all the attacks, this protocol is prone to a threat called denial of service attack. The Signal Stability Table maintains the neighboring node's signal's strength. The authors simulated the protocol in OmNet and a metric known as CPU usage was considered to measure the performance. It proved that when there are malicious nodes the usage of CPU was more than in the absence of malicious nodes.

**2.2. Reasons for Threats and Attacks**

Reasons for threats have been summarized as shown in the following figure4.

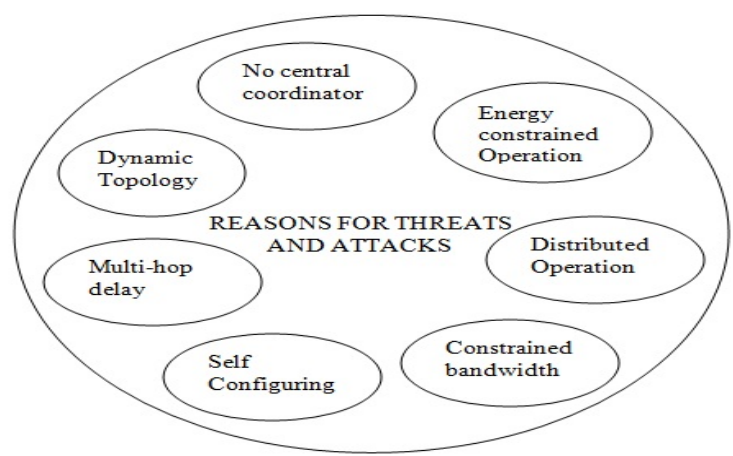

**Figure 4.** Reasons for threats and attacks

For security the authors in [16] have considered the characteristics of the routing protocols in MANETS and conducted a survey on the different types of certificate of authentication to provide better security services. These have been categorized into centralized and distributed certificate authorities. Based on the factors and specifications the appropriate certificate authority can be selected.

## 3. Methodology

The characteristics and approaches to the reactive protocols along with the study of attacks and reasons for such attacks in reactive routing protocols is conducted in order to propose a general framework to avoid such attacks or withstand effectively from the same. The study aims to consider the reactive routing protocols with different types of attacks on them.

The study addresses the following questions:

1. What are the reasons for threats and attacks along with malicious behaviour of the nodes?
2. The after effects of transmitting packets bearing threats and attacks in the network and eliminating them from the network.
3. After the reasons for threats and attacks have been identified, can a generalized framework be proposed to enhance the reactive routing protocols to ensure performance, security and QoS?

DSR (Dynamic Source Routing Protocol) and AODV (Ad Hoc On Demand Distance Vector Routing Protocol) only have been considered here. Analytical Research Methodology has been adopted in conducting the study. The various routing protocols have their own methodologies to send the information to the destination. This research considers the reasons for the threats and attacks in the reactive routing protocols and the effect of eliminating the nodes from the path due to the malicious behavior of nodes. The plan is to propose a strategy which will help in maximizing the throughput and minimizing the routing overhead on the routing protocols and thus help in the selection of the most optimal routing path for any protocol to send the information to the destination.

The research is organized as:
- Study the reactive routing protocols
- Analyze the reasons for threats and attacks in reactive routing protocols

Propose a strategy to handle such behavior to effectively send and receive the information by including the framework in the existing reactive routing protocols.

## 4. Discussions

The study to identify the reasons of threats and attacks showed the weakness of MANETS which put forth the challenges for security in the network with more concentration on how to deal with such threats. The following issues have been identified during the study for reasons of threats and attacks.

1. Nodes as hosts and routers: Because the nodes are playing a dual role of hosts and also as a router in forwarding the packets, if some node becomes a malicious node, there has been a dual disadvantage by misusing the traffic and by dropping the messages.

2. Scarcity of Resources: To make the information secured while transmission by use of cryptographic algorithms makes it complicated to be implemented in an infrastructure less network as the resources are very limited in contrast to the infrastructure network.

3. Mobility: The dynamic changing topology of network gives scope to the faulty nodes more chances for attacks.

While forwarding the packets in the network bearing threats and attacks results in serious problems with sometimes a no possibility to get back to the previous checkpoint having irrecoverable loss. Therefore one of the attacks which show a good behavioral node as malicious for a limited time to steal its identity and be a part of path in route transmission diverts the traffic to a wrong destination. In contrast to this if the nodes that are malicious or tend to be malicious are eliminated from the network, then at some instance there would be very less or no nodes remaining to forward the packets.

When the nodes are cooperating in forwarding and routing these protocols perform better. In this work, the issue of how efficiently can the packets be transmitted in the network considering the fact that there is a type of threat called presence of malicious nodes, has been considered. The malicious behavior of the node due to selfishness or being faulty results in significant performance degradation with more overhead. Due to this increasing overhead, network stability decreases when more nodes become malicious. The increase in malicious nodes reflects an increase of path rejections, so only some percentage of secured paths are available resulting in a better throughput and more overhead.

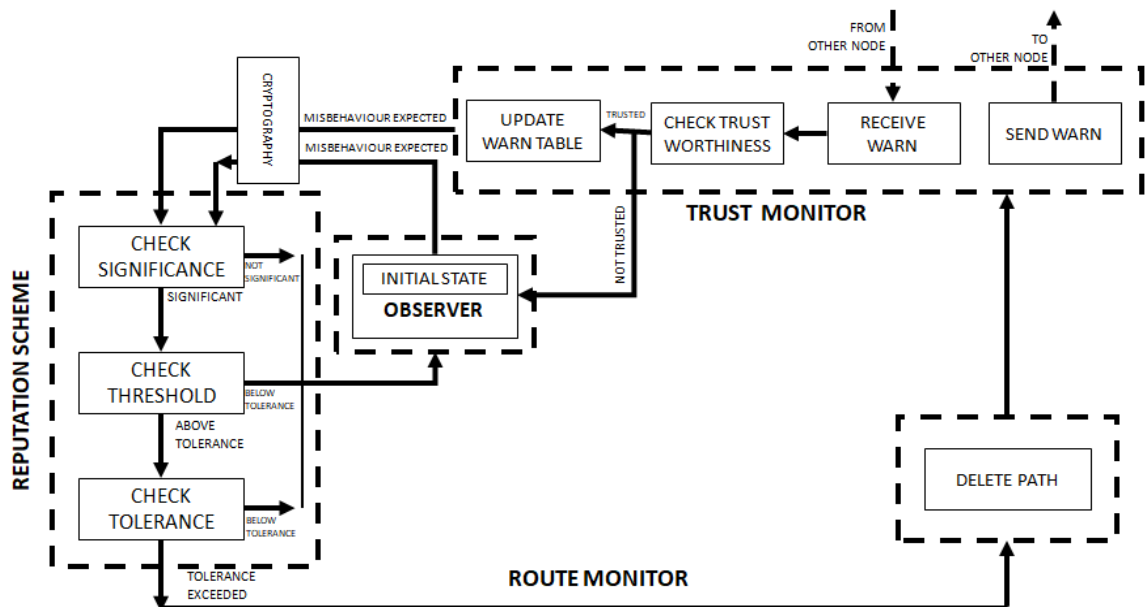

**Figure 5.** Triple factor architecture

In this work, a process of identification of misbehavior node, and action to be taken when it is found is proposed to attain good service without causing interruption to the normal service. Therefore, in this scheme, whenever a faulty node is detected, it triggers a response, i.e., the reaction of neighbor nodes is considered as one of the factors in building the architecture. A node which was mistakenly got into the malicious list or has repented for being malicious in the past turns out to be an ideal node now should be given a chance to integrate back into the network and be a part of routing process. This type of reintegration into the network has been proposed here which increases the paths to the destination and improves the performance.

The proposed triple factor architecture in figure 5, works as follows. Initially when a triggers a warn message then it checks the trust worthiness of such a node and concludes the type of misbehavior. The Reputation scheme verifies the significance of misbehavior. From the reputation rating a node is grouped either as misbehaving or as normal and is based on the severity, that node is eliminated from the routing process if it is proved to be a misbehaving node. The Reputation Scheme manages the accused node information by using Distributed Factor scheme.

In this approach, to attain high security at every node, reputation rating and a trust rating have been included which gives an impression about the neighboring nodes which comes in the communication range. At regular intervals the Direct Factor reputation information is exchanged with others and is considered as the first factor while based on the neighboring nodes information the Trust ratings are updated which is classified as Distributed Factor reputation information with prior reputation ratings is the second factor. And the information to be transmitted uses any basic cryptographic algorithm as the third factor. So this triple factor will help in transmission of packets in a secured way to reach the destination in the reactive routing protocols like DSR and AODV as such protocols allow to integrate this new approach. Moreover DSR and AODV have most common phases of route discovery and route maintenance. Implementation of this phase at every node in route discovery phase will improve the throughput with low overhead on the network in a secured way.

**5. Conclusion and Future Enhancements**

The proposed integrated approach at every node improves the throughput with low network overhead even in presence of malicious nodes. The dynamic behavior of the nodes may turn into malicious or sometimes repent from being malicious. The triple factor inclusion not only allows integrating a faulty misbehavior back into the network after considering the reason for misbehavior but also allows adding basic cryptographic algorithm to have secured packet transmission through the path. As the nodes are not eliminated from the network unless and until they fail in proving their trust, the network is strong and will result in more throughputs with low network overload. Further these protocols can be simulated by making this architecture as a part of the routing process, which will be carried out as our next following research.